\documentclass[a4paper,11pt]{article}
\usepackage{subcaption}
\usepackage{pos}
\usepackage{sidecap}
\usepackage{graphicx}

\usepackage{lineno}
\title{Advancements in the IceAct Energy Spectrum Analysis}

\ShortTitle{IceAct energy spectrum status}

\author{The IceCube Collaboration \\{\normalsize \normalfont(a complete list of authors can be found at the end of the proceedings)}\\}

\emailAdd{larissa.paul@sdsmt.edu}

\abstract{

The IceAct telescopes are Imaging Air Cherenkov telescopes installed as part of the IceCube Neutrino Observatory at the geographic South Pole. They consist of a 61 pixel camera and are small and robust to withstand the harsh environmental conditions. IceAct detects Cherenkov light produced by cosmic-ray particles with energies above approximately 10\,TeV interacting inside the atmosphere, which is complementary to the measurement of the air shower at the surface by IceTop and the high-energy muons in the deep ice. Two telescopes have been taking data since 2019 with a conservative estimated duty cycle of around 10\%. A graph neural network is used to reconstruct the basic air shower properties, like geometry and primary energy. This work focuses on the current progress in analyzing the energy spectrum of cosmic rays using IceAct data.

\vspace{4mm}

{\bfseries Corresponding authors:}
Larissa Paul$^{1*}$, \\

{$^{1}$ \itshape South Dakota School of Mines and Technology}\\
$^*$ Presenter
}
\ConferenceLogo{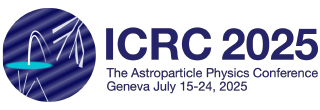}

\FullConference{39th International Cosmic Ray Conference (ICRC2025)\\
 15–24 July 2025\\
Geneva, Switzerland\\}

\begin{document}

\maketitle

\section{Introduction}\label{sec1}
The IceCube Neutrino Observatory located at the geographic South Pole consists of two main components: the in-ice detector, built for the detection of neutrinos, and high-energy muons from cosmic-ray air showers and the surface array, IceTop, built for the detection of cosmic-ray air showers at the surface \cite{IceCube:2019hmk,icetop2013}. Since 2019 the detector setup is expanded by imaging air Cherenkov telescopes called IceAct.\cite{iceact_demonstrator} The telescopes measure the air-shower development in the atmosphere above the other detector components. Measurements of the same air showers in all detector components are a great opportunity for hybrid studies and systematic studies. Additionally, the low-energy threshold of about 10\,TeV of the IceAct telescopes closes the gap between direct and indirect measurements of cosmic-ray particles. 

This work will describe in detail the selection of a dataset for a first physics analysis of IceAct data and the current status of the analysis. For this analysis, two telescopes are used: the roof telescope located above the IceCube Laboratory (ICL) in the center of the IceTop array and the field telescope about 220\,m West of the roof telescope. IceAct telescopes have been taking data continuously since 2019 in different configurations. Since 2021, an all-sky camera has been monitoring the atmosphere above the telescopes, making this year the focus of this work. 
The last part is dedicated to a first attempt to reconstruct the energy of the observed air showers using a graph neural network (GNN). 

\section{Description of the dataset}\label{sec2}
The IceAct data is not yet integrated into the IceCube data stream. Before the data is imported into the IceCube data format, a time synchronization between IceAct and IceCube is performed. The time synchronization allows for pairing the IceAct events with coincident events in the IceCube data stream. Since the detection of air showers with the Cherenkov telescopes depends on atmospheric conditions, to make sure that the detector and atmospheric monitoring cuts are efficient to achieve a homogeneous dataset over time, a 1\% burnsample for IceAct is defined by taking every 100th event in the full dataset. 

Each event is recorded in 61 pixels with 265\,ns long waveforms by the Target DAQ \cite{Author:2017Funktarget}. For each digitized waveform, a pulse extraction is performed. For this, at the highest peak of the waveform, a parabola is calculated for the 3 points around the maximum. If the parabola points downward, meaning the coefficient of the quadratic term is negative, the reconstruction is deemed successful, and the timing and height of the maximum of the parabola are stored as pulse parameters. Additionally, the full width at half maximum is stored. All reconstructed pulses are calibrated from ADC counts to photon-electron(PE) using the calibration constants derived by the pixel calibration method described in \cite{Author:paul2023comp}.

To reduce the number of noise pulses in the images, an image cleaning called  a picture-threshold cleaning is performed. In short, a picture threshold is defined above which the signal in the pixel is clearly photon-induced. All pixels adjacent to at least two pixels above the picture-threshold are also considered photon-induced signals if their pulse height is above a defined boundary threshold and within a certain time window. Additionally, all pixels need to form groups of at least 3 neighboring pixels for the image to be considered in further analysis. 

After the image cleaning, a containment cut is applied. The image is kept if the sum of the pixel heights on the ring of all outer pixels on the camera edge are equal or less than the sum of pixel heights of the inner pixels.

\section{Environmental monitoring}\label{sec3}
An all-sky camera continuously monitors the night sky to track atmospheric conditions during our measurements.
It is situated on top of the ICL.
Since 2021, during our data-taking season, we regularly take images with 3 different exposures (20\,s, 100\,s, 180\,s). A photo with the same exposure is taken roughly every 8 minutes. The resolution of the images is 640 times 480 pixels. 

\begin{figure}[]
  \centering
  \begin{minipage}[t]{0.45\textwidth}
    \includegraphics[width=\textwidth]{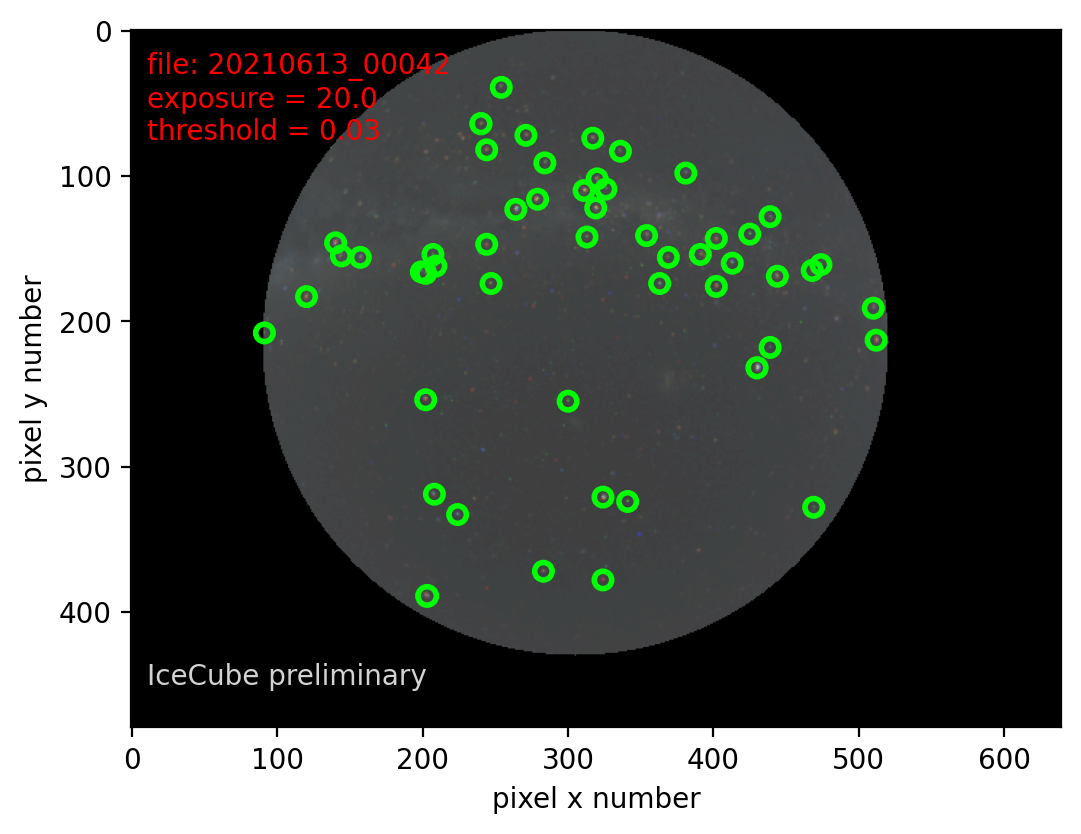}
    \caption{Example image of the all-sky camera data on a clear day. The Milky Way is in the background and the green circles indicate positions for identified stars.}\label{fig:all-sky camera_clear}
  \end{minipage}
  \hfill
  \begin{minipage}[t]{0.52\textwidth}
    \includegraphics[width=\textwidth]{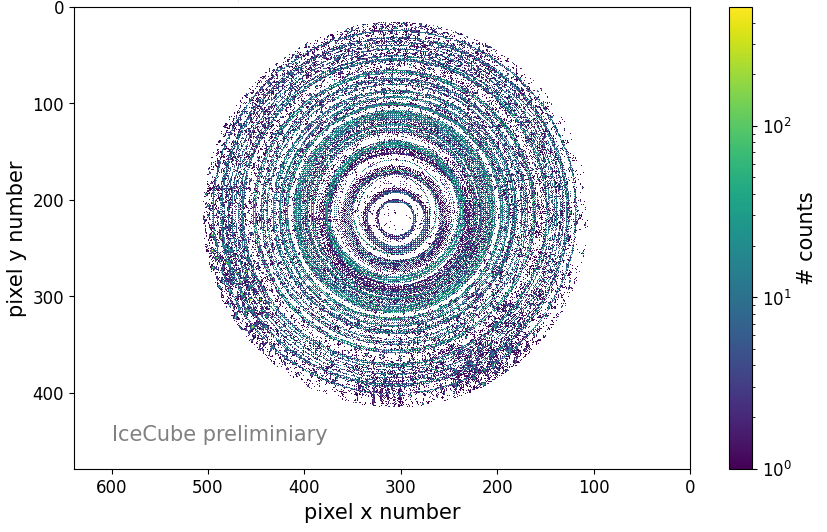}
    \caption{Result of the full season analysis tracking the stars. Clearly visible are the circular tracks resulting from the rotation of the Earth.}\label{fig:startracking}
  \end{minipage}
\end{figure}

A standard library is used to find the stars in all-sky camera images \cite{opencv_library}. An examples of all-sky camera image is shown in Figure \ref{fig:all-sky camera_clear}. The image is taken during clear conditions. Even during very high aurora activity the star finding algorithm works. Figure \ref{fig:startracking} shows all the stars found during the whole data taking season and their circular tracks created due to the Earth's rotation.

\begin{figure}
    \centering
    \includegraphics[width=0.9\linewidth]{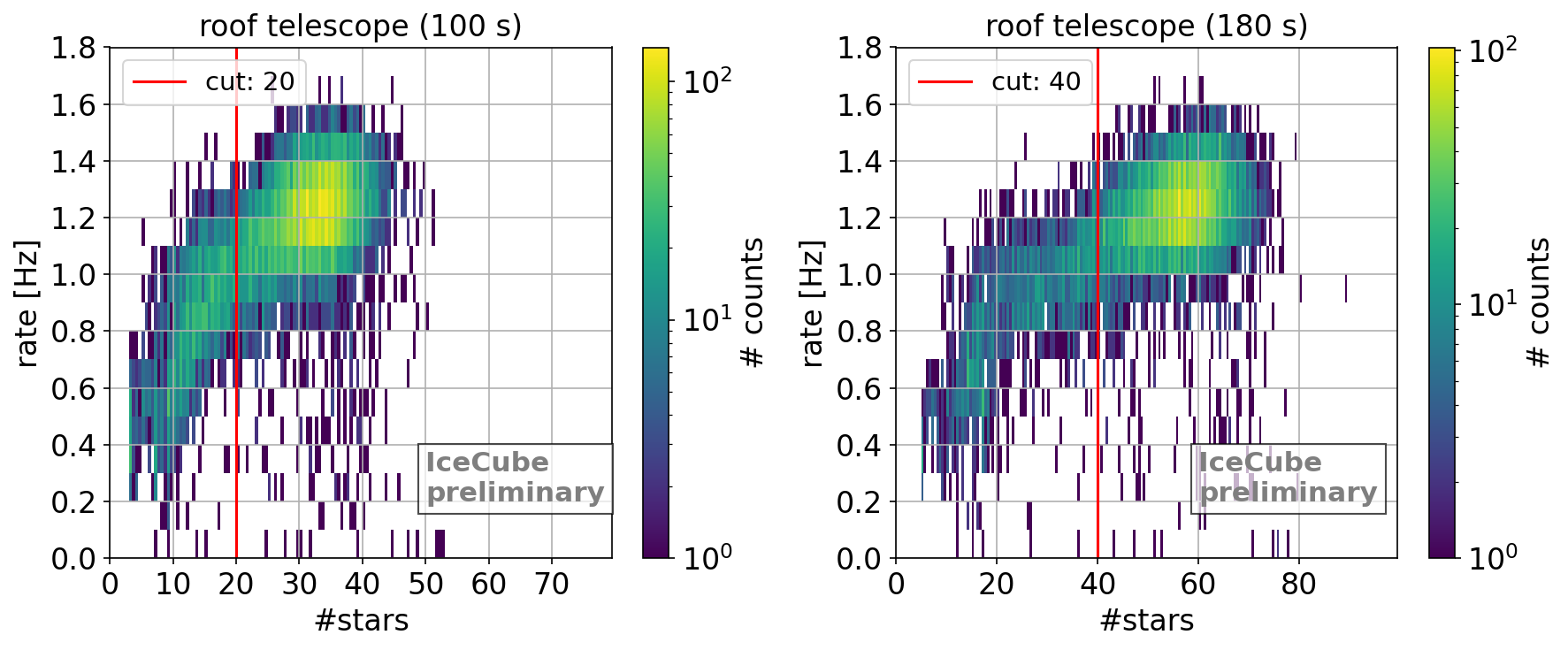}
    \caption{Histogrammed data of the event rate of one telescope versus the number of stars detected in an all-sky camera image. The left plot shows the data for an exposure of 100\,s, the right plot shows the 180\,s exposure. Depending on the exposure 100\,s/180\,s the event rate of the telescopes drops when less than 20/40 stars are visible in the all-sky camera images.}
    \label{fig:ratevsstars}
\end{figure}

To determine the standard number of stars visible during optimal operations we correlate the number of detected stars to the rate measured by the telescopes. For this, very strict image cleaning parameters are used to achieve a telescope rate dominated by cosmic-ray air showers. The correlation between the event rate of a telescope and the number of stars visible in an all-sky camera image is shown in Figure \ref{fig:ratevsstars}. Depending on the exposure of the images, the rate of the telescope is dropping if the number of stars is below 20(40) for an exposure of 100\,s(180\,s), indicating that clouds cover at least part of the sky, making the data taking period not suitable for cosmic-ray analysis. Therefore, those time periods are excluded.

\section{Additional data quality cuts}\label{sec4}
The telescopes operate fully autonomously. If the light conditions change due to the moon or auroras, some detector settings change automatically, which can have an influence on the measurement. The changing properties are, for example, 
 the trigger threshold to reduce the noise of the brighter sky and/or the voltage supplied to the silicon photomultiplier (SiPM) if the sky gets so bright that adjusting the threshold does not reduce the trigger rate efficiently.
 To prevent systematic effects from these non-standard settings from impacting the analysis, we exclude time periods when the detector operations settings are not the standard setting.

In addition to the number of stars, a green value is calculated to determine the current aurora activity. The green value is the sum of the green values of the image in an RGB-format. We reject all events if the sum of the green values in the image is above 2.8e6(4e6) for the 100\,s(180\,s) exposure.

To improve stability for this first analysis, we exclude all time periods for which the sun is close to the horizon
or the moon is above the horizon and while the moon phase is more than 30\%. 

\begin{figure}
    \centering
    \includegraphics[width=0.98\linewidth]{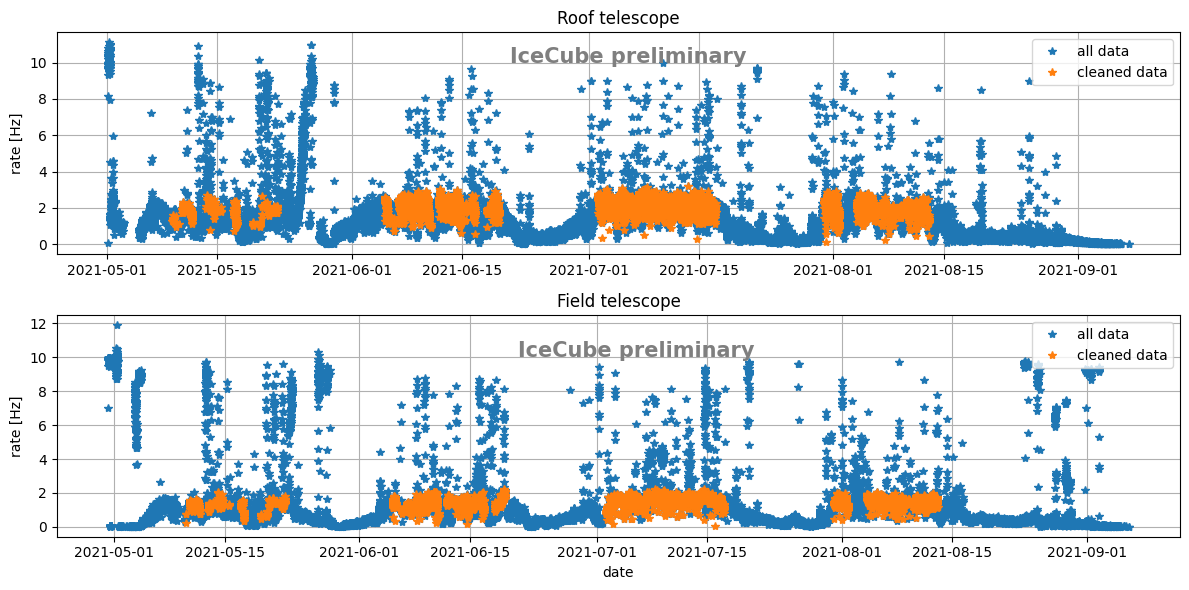}
    \caption{Plotted are the event rate for the two telescopes versus time. The top panel shows the roof telesocpe data, the bottom panel shows the field telescope data. After applying all detector-related cuts, the event rate in the telescopes seems to be stable over time.}
    \label{fig:ratevsyear}
\end{figure}

Figure \ref{fig:ratevsyear} shows the raw trigger rate of the detector versus time. The overlayed orange points show the rate after all above-described cuts are applied. The rate curve is more stable and indicates an appropriate selection of good data-taking periods. In total we have about 22\,days of good observation time for the selection criteria we defined above. This can possibly be further improved by taking a detailed look at the currently excluded time periods. The final 1\% burnsample consist of 20456 events for the roof telescope and 15511 events for the field telescope. The slightly different camera components are responsible for a lower trigger rate of the field telescope resulting in less events in the final event sample.

\section{Simulation}\label{sec3}
A description of the Monte Carlo (MC) simulation dataset can be found in \cite{Author:paul2023comp}. It consists of 5 primary nuclei (proton, Helium, Nitrogen, Aluminum and Iron). The showers are simulated around the center point between the roof and the field telescope. The radius starts from 250\,m between an energy of 
 3 to 4 in log(E/GeV) and increases with each quarter decade in energy by 50\,m. Above 6\,PeV, the MC statistics drop, and no showers above an energies of 6.75\,PeV have been simulated.  

\begin{figure}
    \centering
    \includegraphics[width=0.98\linewidth]{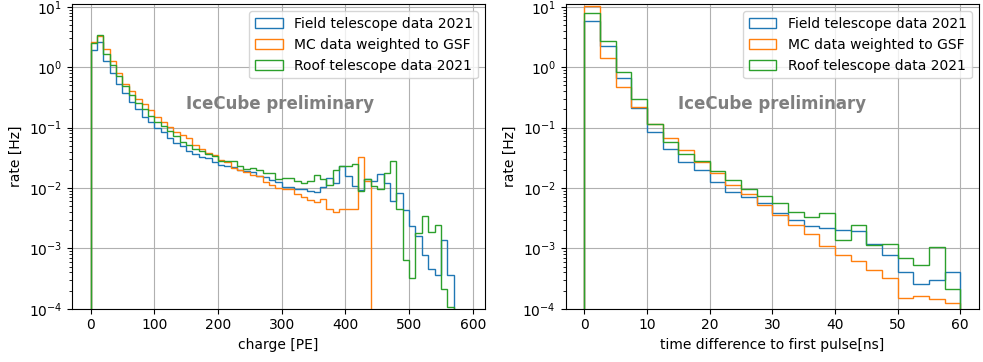}
    \caption{A comparison of height (left) and time (right) of the reconstructed pulses, for the data and the MC simulation. There is a mismatch at the higher end of the distributions due to pixel-to-pixel variation. Additionally, the MC set above 1\,PeV has quite low statistics, and we do not simulate events above 6.75\,PE, which might lead to mismatches for the higher energies.}
    \label{fig:datamc}
\end{figure}

A comparison of the reconstructed pulse heights and times of the burnsample data with the MC is shown in Figure \ref{fig:datamc}.The simulation data in all the plot is weighted to the Global Spline Fit (GSF) model \cite{Dembinski:2017N7}. There is a mismatch at the higher end of the distributions due to pixel-to-pixel variations in the data, which have not yet been accounted for in the analysis. This will be improved before a full spectrum analysis. Additionally, the MC set above 1\,PeV has quite low statistics, and we do not simulate events above 6.75\,PeV, which might lead to mismatches at the higher energies.

Future work will also include an attempt to recover any saturated signals. 
Pulse shaping ensures that even saturated signal heights can be recovered, thanks to a trailing shoulder in the waveform.\cite{Author:2017Funktarget}
The length of the shoulder corresponds to the unsaturated signal height.

\section{GNN results}\label{sec4}
A graph neural network (GNN) is used to reconstruct the shower properties. A description of the GNN structure can be found in \cite{Author:paul2023comp}. The reconstruction uses a single telescope. The input into the network per pixel is a unique pixel id, the signal height, and the timing of the signal relative to the first pulse in the image. The GNN output parameters are the total energy of the air shower, the shower core position, the shower direction and the depth of the shower maximum.
The shower core is reconstructed by determining its distance from the telescope and the opening angle $\theta$ between the x-axis and the direction of the shower core position. To avoid the circular symmetry for theta we reconstruct both the sin($\theta$) and cos($\theta$). Similarly we encode the shower direction in zenith, cos(azimuth) and sin(azimuth). 

\begin{SCfigure}
    \centering
    \includegraphics[width=0.5\linewidth]{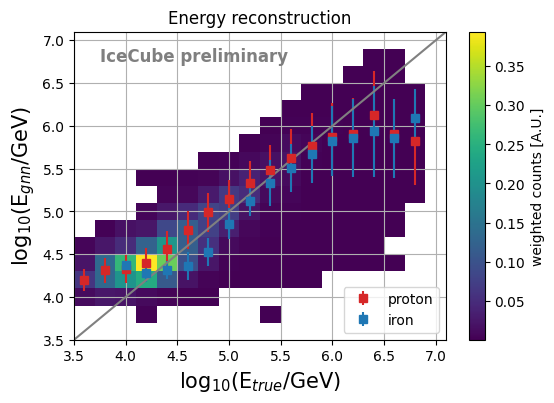}
    \caption{Reconstructed energy vs. true energy for the graph neural network reconstruction. In the region with high MC statistics, the correlation is linear. There is a composition bias for the lowest energies.}
    \label{fig:energyresolution}
\end{SCfigure}

Figure \ref{fig:energyresolution} shows the reconstructed versus the true simulated energy. Due to trigger threshold effects, below log(E/GeV)=4.5 of true energy, the reconstructed energy is biased and nearly flat. For the energy up to about log(E/GeV)=5.75 the reconstruction is to the identity line created by the true energy. Due to lower statistics at energies above log(E/GeV)=5.75, the reconstruction also gets biased again to smaller reconstructed energies than the true energy. There is a small composition bias for the lower energies, which seems to decrease for the higher energies. 

\section{Burn sample results}\label{sec5}
\begin{figure}
    \centering
    \includegraphics[width=0.9\linewidth]{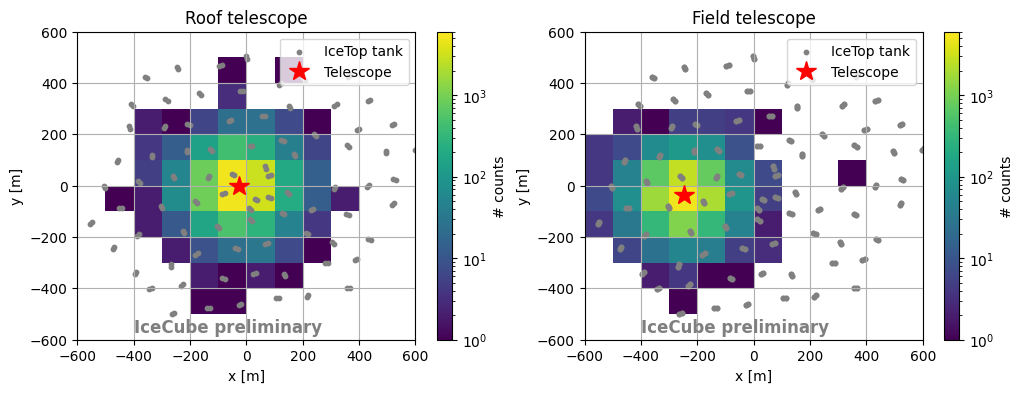}
    \caption{Reconstructed shower core positions around roof (left) and field (right) telescope. As expected the event rate decreases with the distance from the telescope.}
    \label{fig:xy_position}
\end{figure}

The trained GNN is applied to the 1\% burnsample after cuts. Looking at the reconstructed shower core positions in Figure \ref{fig:xy_position}, the distribution seems the reconstruction rate decreases with the radius because of the energy dependence of the light emission. For larger distances, only higher energy showers produce enough light to trigger the telescope.

\begin{figure}
    \centering
    \includegraphics[width=0.99\linewidth]{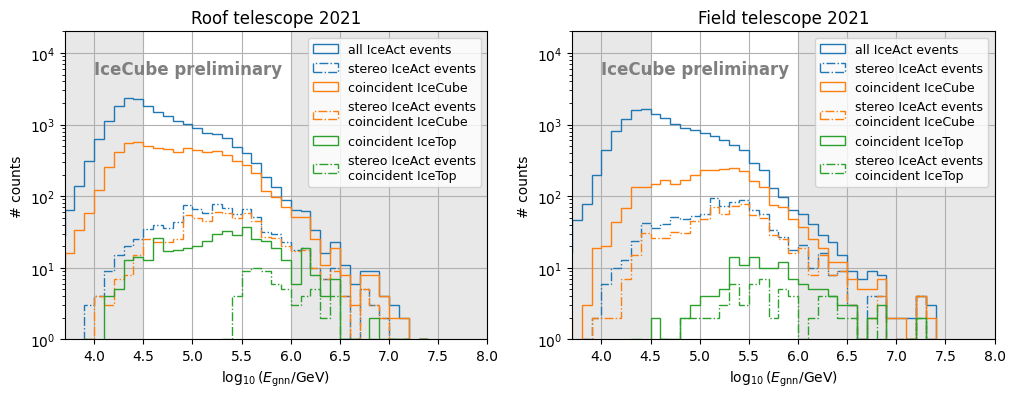}
    \caption{Reconstructed energies for both telescopes: the roof telescope data on the left and the field telescope data on the right.  The event count decreases with higher energies as expected. There is a steeper decline at around log(E/GeV)=5.5, which is related to the flattening of the reconstructed energy versus the true energy.
    Grey areas are either below the full efficiency of the telescope in this data selection or not well covered by the MC data sample. For both areas the GNN is also not performing well.}
    \label{fig:rates}
\end{figure}

Figure \ref{fig:rates} shows the event count as a function of the reconstructed energies. The curves show a falling spectrum as expected. The different colors indicate different event selections. Blue are all IceAct events, orange are IceAct events which have a coincident IceCube in-ice event, green are events with a coincident IceTop event. The higher energy threshold for the IceCube in-ice detector and the IceTop array is clearly visible. The low-energy peak of the IceCube events in the roof telescope data, which is not present in the field telescope data,
is due to an integrated trigger. The trigger of the roof telescope triggers a read out of the full IceCube detector including the IceTop array, which keeps all triggered events in the IceCube data stream. For the field telescope, the IceCube events need to pass a filter to be synchronized with IceAct data. The low-energy part of the coincident IceTop events in the roof telescope rates originates in the IceTop InFill, a more densely instrumented part of the array, which is close to that telescope \cite{icetop2013}.  

The dashed lines are events that are present in both telescopes datasets. As expected the distributions for the dashed lines are shifted to higher energies with respect to the single telescope distribution, which is expected since events detected by both telescopes should have on average a higher energy. Since this is a single telescope reconstruction, the distributions for both telescopes do not match exactly, but it shows nicely that only events with high energies are seen in both telescopes.

\begin{SCfigure}
    \centering
    \includegraphics[width=0.65\linewidth]{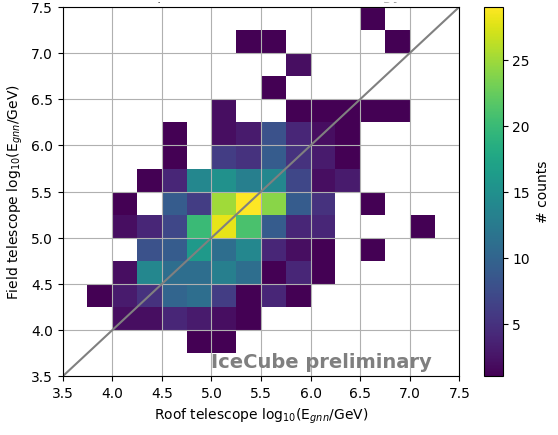}
    \caption{Correlation of reconstructed energies for events that are present for both telescopes. The reconstructed energies are linearly correlated, indicating that the telescopes are measuring the same event and that the energy reconstruction works.}
    \label{fig:energy_correlation}
\end{SCfigure}

Figure \ref{fig:energy_correlation} shows the correlation between the reconstructed energies in the two telescopes.
There seems to be a tendency that energies in the roof telescopes are reconstructed lower than the reconstruction in the field telescope which is under investigation.

\section{Summary and conclusion}\label{sec6}
We successfully identified parameters to create a high-quality dataset for a cosmic-ray analysis, resulting in an observation time of about 22 days. For future analysis, some possible improvements could be achieved by further studying the currently excluded time periods.

A GNN reconstruction was successfully applied on a 1\% burnsample of the observation time. The performance of the analysis is quite promising for a single telescope reconstruction. The energy reconstruction generally seems to work well. There are several improvements that will be applied before the final spectrum analysis, including accounting for pixel-to-pixel variations and refining the treatment of events with saturated pulses.
Additionally, including auxiliary network inputs like IceCube and/or IceTop parameters when available while training the GNN should improve the results further. Another approach could be to train the events simultaneously on images of both telescopes when available. Finally, some minor improvements could made by tuning the hyperparameters and network structure of the GNN.

\bibliographystyle{ICRC}
\bibliography{references}

\clearpage

\section*{Full Author List: IceCube Collaboration}

\scriptsize
\noindent
R. Abbasi$^{16}$,
M. Ackermann$^{63}$,
J. Adams$^{17}$,
S. K. Agarwalla$^{39,\: {\rm a}}$,
J. A. Aguilar$^{10}$,
M. Ahlers$^{21}$,
J.M. Alameddine$^{22}$,
S. Ali$^{35}$,
N. M. Amin$^{43}$,
K. Andeen$^{41}$
% ...
\\
\\
% ...
$^{10}$ Universit{\'e} Libre de Bruxelles, Science Faculty CP230, B-1050 Brussels, Belgium \\
% ...
$^{16}$ Department of Physics, Loyola University Chicago, Chicago, IL 60660, USA \\
$^{17}$ Dept. of Physics and Astronomy, University of Canterbury, Private Bag 4800, Christchurch, New Zealand \\
% ...
$^{21}$ Niels Bohr Institute, University of Copenhagen, DK-2100 Copenhagen, Denmark \\
$^{22}$ Dept. of Physics, TU Dortmund University, D-44221 Dortmund, Germany \\
% ...
$^{35}$ Dept. of Physics and Astronomy, University of Kansas, Lawrence, KS 66045, USA \\
% ...
$^{39}$ Dept. of Physics and Wisconsin IceCube Particle Astrophysics Center, University of Wisconsin{\textemdash}Madison, Madison, WI 53706, USA \\
% ...
$^{41}$ Department of Physics, Marquette University, Milwaukee, WI 53201, USA \\
% ...
$^{43}$ Bartol Research Institute and Dept. of Physics and Astronomy, University of Delaware, Newark, DE 19716, USA \\
% ...
$^{63}$ Deutsches Elektronen-Synchrotron DESY, Platanenallee 6, D-15738 Zeuthen, Germany \\
$^{\rm a}$ also at Institute of Physics, Sachivalaya Marg, Sainik School Post, Bhubaneswar 751005, India \\
% ...

\subsection*{Acknowledgments}

\noindent
The authors gratefully acknowledge the support from the following agencies and institutions:
USA {\textendash} U.S. National Science Foundation-Office of Polar Programs,
U.S. National Science Foundation-Physics Division,
U.S. National Science Foundation-EPSCoR,
U.S. National Science Foundation-Office of Advanced Cyberinfrastructure,
Wisconsin Alumni Research Foundation,
Center for High Throughput Computing (CHTC) at the University of Wisconsin{\textendash}Madison,
Open Science Grid (OSG),
Partnership to Advance Throughput Computing (PATh),
Advanced Cyberinfrastructure Coordination Ecosystem: Services {\&} Support (ACCESS),
Frontera and Ranch computing project at the Texas Advanced Computing Center,
U.S. Department of Energy-National Energy Research Scientific Computing Center,
Particle astrophysics research computing center at the University of Maryland,
Institute for Cyber-Enabled Research at Michigan State University,
Astroparticle physics computational facility at Marquette University,
NVIDIA Corporation,
and Google Cloud Platform;
Belgium {\textendash} Funds for Scientific Research (FRS-FNRS and FWO),
FWO Odysseus and Big Science programmes,
and Belgian Federal Science Policy Office (Belspo);
Germany {\textendash} Bundesministerium f{\"u}r Forschung, Technologie und Raumfahrt (BMFTR),
Deutsche Forschungsgemeinschaft (DFG),
Helmholtz Alliance for Astroparticle Physics (HAP),
Initiative and Networking Fund of the Helmholtz Association,
Deutsches Elektronen Synchrotron (DESY),
and High Performance Computing cluster of the RWTH Aachen;
Sweden {\textendash} Swedish Research Council,
Swedish Polar Research Secretariat,
Swedish National Infrastructure for Computing (SNIC),
and Knut and Alice Wallenberg Foundation;
European Union {\textendash} EGI Advanced Computing for research;
Australia {\textendash} Australian Research Council;
Canada {\textendash} Natural Sciences and Engineering Research Council of Canada,
Calcul Qu{\'e}bec, Compute Ontario, Canada Foundation for Innovation, WestGrid, and Digital Research Alliance of Canada;
Denmark {\textendash} Villum Fonden, Carlsberg Foundation, and European Commission;
New Zealand {\textendash} Marsden Fund;
Japan {\textendash} Japan Society for Promotion of Science (JSPS)
and Institute for Global Prominent Research (IGPR) of Chiba University;
Korea {\textendash} National Research Foundation of Korea (NRF);
Switzerland {\textendash} Swiss National Science Foundation (SNSF).

\end{document}